\documentclass[fleqn,12pt,twoside]{article}
\usepackage{amsmath}
\usepackage{amssymb}
\usepackage{espcrc1}

\usepackage{graphicx}
\usepackage[figuresright]{rotating}

\newcommand{\ds}{\displaystyle}

\newcommand{\nb}{\nonumber}

\newcommand{\be}{\begin{equation}}
\newcommand{\ee}{\end{equation}}
\newcommand{\bqr}{\begin{eqnarray}}
\newcommand{\eqr}{\end{eqnarray}}
\newcommand{\bc}{\begin{center}}
\newcommand{\ec}{\end{center}}
\newcommand{\efc}{\end{figure}\ec}
\newcommand{\etal}{{\it et~al.}}
\newcommand{\demi}{\frac{1}{2}}
\newcommand{\thuit}{\frac{3}{8}}
\newcommand{\rhoi}{\rho^{i/3}}
\newcommand{\rhoj}{\rho^{j/3}}
\newcommand{\rhok}{\rho^{k/3}}
\newcommand{\Ps}{P_{\sigma}}
\newcommand{\dd}{\delta(\hbox{\bf r})}
\newcommand{\bsigma}{\mbox{\boldmath $\sigma$ \unboldmath}}

\newcommand{\br}{\hbox{\bf r}}
\newcommand{\PP}{\hbox{\bf P}}
\newcommand{\PD}{\hbox{\bf P}^2}
\newcommand{\PPP}{\hbox{\bf P}^{\, '}}
\newcommand{\PPD}{\hbox{\bf P}^{\, ' 2}}
\newcommand{\RR}{\hbox{\bf R}}
\newcommand{\tpid}{\left(\frac{3\pi^2}{2}\right)^{2/3}}
\newcommand{\rdt}{\rho_0^{2/3}}
\newcommand{\rct}{\rho_0^{5/3}}
\title{Compressibility, effective mass and density dependence in Skyrme forces}

\author{B. Cochet\address[IPNL]{IPN Lyon,
CNRS-IN2P3/UCB Lyon 1, B\^at. Paul Dirac, \\
        43, Bd. 11.11.18, 69622 Villeurbanne Cedex, France}%
        \thanks{Email: {\tt cochet@ipnl.in2p3.fr}.},
        K. Bennaceur\addressmark[IPNL]
        \thanks{Email: {\tt bennaceur@ipnl.in2p3.fr}.},
        P. Bonche\address[SPHT]{Service de Physique Th\'eorique,
        CEA Saclay, 91191 Gif-sur-Yvette Cedex, France},
        T. Duguet\addressmark[SPHT]
        \thanks{present address: Argonne National Lab.,
        Physics Division, 9700 S. Cass Av., Argonne, Ill., 60439, USA},
        and
        J. Meyer\addressmark[IPNL]\thanks{Email: {\tt jmeyer@ipnl.in2p3.fr}.}}

\begin{document}

\maketitle

\begin{abstract}
Generalized density dependence in Skyrme effective interactions is investigated
to get forces valid beyond the mean field approximation.
Preliminary results are presented for infinite symmetric
and asymmetric nuclear matter up to pure neutron matter.  
\end{abstract}

\section{INTRODUCTION}

It is commonly accepted that it does exist a relation between 
compressibility, effective mass and the density dependence of a given 
effective force.
Studies with Skyrme forces~\cite{sly4} have shown that the incompressibility 
$K_\infty$ and the effective mass cannot be chosen independently
once the analytical form of the (single) density-dependent term has been
chosen.
This has led to the $\rho^{1/6}$ density dependence in Skyrme forces
like SkM$^*$ which allows a value of $K_\infty$ around $220$~MeV 
close to that extracted from the experimental breathing mode 
analyses~\cite{blaizot_rep,blaizot_95} and an effective mass $m^*$ around 
$0.7\,m$ simultaneously.  The present status about
compressibility has been summarized by G.~Col\`o during this
conference~\cite{cologiai}.

Recently, the density dependence of phenomenological effective interactions,
such as Skyrme or Gogny forces, has been revisited in the context of beyond 
mean field calculations~\cite{duguet1,duguet2}. 
Indeed, while a dependence of the interaction on the density 
is well established for calculations at the mean field level~\cite{NV72}, 
no strongly motivated prescription exists when several mean fields are 
mixed as in the Generator Coordinate Method (GCM) and the 
Projected Mean Field Method (PMFM).
First, an extension of the Goldstone-Brueckner theory has motivated 
the GCM and the PMFM from a perturbative point of view for the first 
time~\cite{duguet1}. 
In this extended context, a generalized Brueckner $G$ matrix summing
particle-particle ladders has been defined and may be used as a reference 
from which phenomenological interactions in GCM or PMFM calculations
should be approximated.
It is possible to simplify this in-medium interaction to 
extend the validity of the Skyrme force~\cite{skyrme,VB72} and to 
identify approximately the density dependence originating from Brueckner 
correlations in the context of mixed nonorthogonal vacua~\cite{duguet2}. 
The renormalization of three-body forces has also been advocated
as an important source of density dependence in phenomenological 
interactions. 
Indeed, their renormalization through a density-dependent two-body interaction 
has been proved to be justified at the mean field level in some particular 
cases~\cite{NV72,VB72}. 
As for the resummation of Brueckner correlations, the corresponding density 
turns out to be the one of the mean field which is calculated.
Consequently, a single density-dependent term has often been used in 
phenomenological forces, e.g. Gogny~\cite{decharge} and Skyrme~\cite{VB72} 
effective interactions. 
On the other hand, the density dependence renormalizing three-body force
effects has
been shown to be different from the one taking care of Brueckner correlations
when going beyond the mean field approximation~\cite{duguet2}.
In fact, the use of two different, and theoretically motivated,
density-dependent
terms happens to have non negligible effects on collective
spectra~\cite{thomas_these}.
Based on this analysis, it is legitimate to redefine the Skyrme
interaction at the mean field level.
Including two density-dependent terms, each related to a given physical
origin,  will allow a proper extrapolation of the interaction to GCM and
PMFM calculations.

A first attempt is presented in this paper, where
we concentrate on the practical
advantages of having two different density-dependent terms.
Relating them to the resummation of Brueckner correlations and to the
renormalization of three-body forces is the aim of our on-going work.
In this way, this paper is concerned by a systematic investigation
of infinite matter, symmetric and asymmetric up to pure neutron matter,
using multi density-dependent terms in Skyrme forces.
Our analysis is focused on $\rho^{1/3}$ terms which are connected with
a $k_F$ expansion of the Brueckner $G$ matrix.

\section{STANDARD SKYRME EFFECTIVE INTERACTIONS}
\label{sec:standard}

\begin{figure}[htb]
\bc
\includegraphics*[bb= 124 462 426 675,clip,scale=1.05]{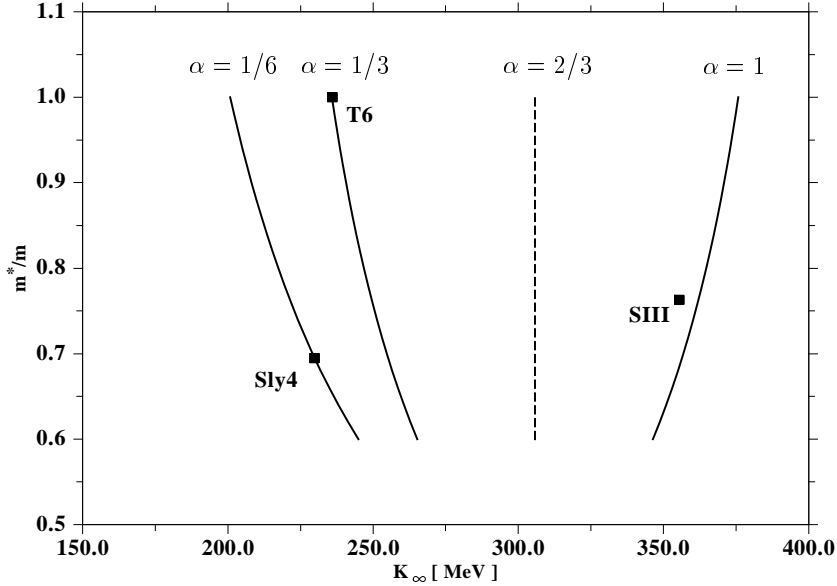}
\ec
\vskip -1cm
\caption{Correlation between incompressibility, 
effective mass and density dependence of standard parameterizations 
of Skyrme effective interactions.}
\label{fig:corrKm}
\end{figure}

When using Skyrme effective forces with a standard density dependence
$\rho^\alpha$, where $\rho$ is the total density, the adjustable
parameter $\alpha$ is strongly correlated to the
incompressibility and the effective mass in nuclear matter.
This is illustrated on Figure~\ref{fig:corrKm} for different standard 
Skyrme forces with various powers $\alpha$. 
In all cases, the different parameters of the force have been adjusted 
in order to obtain the saturation density $\rho_0=0.16$~fm$^{-3}$ and 
the energy per particle $E/A=-16$~MeV in infinite symmetric nuclear matter.
Once these two properties are fixed, the relation
between $m^*/m$ and $K_\infty$ is entirely determined by $\alpha$.
This well-known feature, discussed in~\cite{sly4}, yields to the 
conclusion that a value of $\alpha$ around 1, like in SIII, does not allow 
to reach a correct compression modulus~\cite{blaizot_rep,giant}.
Only values of $\alpha$ ranging from $1/6$ to $1/3$ allow for an
acceptable set $\{m^*/m,K_\infty\}$.

%

It is enlightening to study a bit further the relation  between
$m^*/m$ and $K_\infty$ in the case of the standard parameterizations
of Skyrme forces. 
Introducing as in~\cite{sly4} the notation $\Theta_s=3t_1+(5+4x_2)t_2$,
one has
\begin{equation}
\frac{m^*}{m}=\left(1+\frac{1}{8}\frac{m}{\hbar^2}\rho_0\Theta_s
\right)^{-1},
\label{eq:effmass}
\end{equation}
and
\begin{equation}
K_\infty=-\frac{3\hbar^2}{5m}\left(\frac{3\pi^2}{2}\right)^{2/3}
\rho_0^{2/3}+\frac{3}{8}\Theta_s\left(\frac{3\pi^2}{2}\right)^{2/3}
\rho_0^{5/3}
+\frac{9}{16}\alpha(\alpha+1)t_3\rho_0^{\alpha+1}.
\label{eq:kinf}
\end{equation}
The energy per particle in infinite nuclear matter at saturation
density is given by:
\begin{equation}
\frac{E}{A}=\frac{3\hbar^2}{10m}\left(\frac{3\pi^2}{2}\right)^{2/3}
\rho_0^{2/3}
+\frac{3}{8}t_0\rho_0
+\frac{3}{80}\Theta_s\left(\frac{3\pi^2}{2}\right)^{2/3}\rho_0^{5/3}
+\frac{1}{16}t_3\rho_0^{\alpha+1}.
\label{eq:ea}
\end{equation}
The saturation density being determined by the condition on the
pressure $P(\rho_0)=0$ which gives (if $\rho_0\neq0$)
\begin{equation}
\frac{\hbar^2}{5m}\left(\frac{3\pi^2}{2}\right)^{2/3}\rho_0^{2/3}
+\frac{3}{8}t_0\rho_0
+\frac{1}{16}\Theta_s\left(\frac{3\pi^2}{2}\right)^{2/3}\rho_0^{5/3}
+\frac{1}{16}t_3(\alpha+1)\rho_0^{\alpha+1}=0\,.
\label{eq:pressure}
\end{equation}
The three equations~(\ref{eq:kinf}), (\ref{eq:ea}) and~(\ref{eq:pressure})
form a system with three unknown quantities
$t_0$, $t_3$ and $\Theta_s$. Solving this system for a given set
$(\rho_0,E/A,K_\infty)$ provides the coefficients $t_0$, $t_3$
and $\Theta_s$, this latter determining uniquely the effective
mass $m^*/m$ through~Eq.~(\ref{eq:effmass}).

Actually, the previous conclusion is incorrect in one special case.
To see that, let us set
\begin{equation}
F(\Theta_s,t_3)=\frac{1}{16}
\left[\Theta_s\left(\frac{3\pi^2}{2}\right)^{2/3}
+\frac{5}{3}\,t_3\right]\,.
\label{eq:f}
\end{equation}
It is trivial to check that if and only if $\alpha=2/3$ we have

\be
\left\{
  \begin{matrix}
    \ds
    - \frac{3\hbar^2}{5m}\tpid \rdt &&& 
    + & \ds 6\,F(\Theta_s,t_3)\rct & = & K_\infty\,, \\[3mm]
    \ds
    \frac{3\hbar^2}{10m}\tpid \rdt & + & \ds \thuit t_0\rho_0 & 
    + & \ds \frac{3}{5}\,F(\Theta_s,t_3)\rct & = & \ds\frac{E}{A}\,, \\[3mm]
    \ds
    \frac{\hbar^2}{5m}\tpid \rdt & + & \ds \thuit t_0\rho_0 &
    + & F(\Theta_s,t_3) \rct & = & 0\,. \\[2mm]
  \end{matrix}
\right.  
\label{eq:sys}
\end{equation}

In this particular case, there are three equations and two unknown
quantities, $t_0$ and $F(\Theta_s,t_3)$, and, in general, this system
has no solution.
However, we can use the last two equations to determine $t_0$
and $F(\Theta_s,t_3)$
and then calculate the corresponding value of $K_\infty$.
Because of the definition (\ref{eq:f}), $\Theta_s$ can be freely chosen.
This situation is illustrated on Figure~\ref{fig:corrKm}
by the vertical dashed line, when $\alpha=2/3$, the value
of $K_\infty$ is fixed and the effective mass can be freely chosen.
However,
this case is not interesting since it does not correspond to a
realistic value of $K_\infty$.

%
%

\section{MODIFIED DENSITY DEPENDENCE IN SKYRME FORCES}

Following several previous attempts~\cite{farine,mike} and adding a density 
dependence on each term of the force, one can investigate the properties
of generalized Skyrme effective interactions written as:

\bqr
V(\br_1,\br_2) & = & \sum_i \, t_{0i} \, \rhoi \, (1+x_{0i}\Ps) \, \dd \nb\\ 
               & + & \demi \, \sum_j \, t_{1j} \, \rhoj
                           \, (1+x_{1j}\Ps) \,
                 \left[\dd \PPD + \PD \dd\right] \nb \\ 
               & + &\sum_j \, t_{2j} \, \rhoj 
                          \, (1+x_{2j}\Ps) \PPP \cdot \dd \PP \\
               & + & i \sum_k \, W_{0k} \, \rhok \, \bsigma \cdot 
                              \left[\PPP \times \dd \PP\right] \nb \\
\nb                                      
\label{eq:sky_full}
\eqr
where $\rho \equiv \rho(\RR)$ is the total density and see ref.~\cite{sly4} 
for the other notations.
Standard parameterizations discussed in Section~\ref{sec:standard}
correspond to $j=k=0$ and $i=0,3$ in SIII and $i=0,1/2$ in SkM$^*$ and 
SLy4 for instance.

This generalized parameterization increases the number of
parameters in the force, making it more phenomenological and reducing
its predictive power. Besides, the fitting procedure may become
intractable if most of the parameters are free. In this first exploratory
work, we limit our study to the case with
$t_{1j}=t_{2j}=0$ except for $j=0$ (hereafter we omit the second
index) and three
non zero values for $t_{0i}$, namely $i_1$, $i_2$ and $i_3$ out of the set
$\{0,1,2,3\}$. The corresponding forces will be labeled by
$[i_1,i_2,i_3]$. The spin-orbit term will not be discussed here since
it has no contribution in infinite nuclear matter.

The problems faced with the Skyrme forces with one density-dependent
term are cured with this family of parameterizations.
Following the same procedure to adjust the coefficients, we can fix
the saturation properties of infinite symmetric nuclear matter
$\rho_0$, $E/A$, $K_\infty$ and $m^*/m$ independently and determine the
coefficients $t_{0i}$ of the force.

However this procedure only guarantees to have realistic
properties of nuclear matter in the vicinity of the saturation
point. If we want a force which gives an equation of state
in agreement with {\it ab initio} calculations at low
density as well as high density (up~to~$\sim\,3\rho_0$)
one should adopt a different fitting procedure. First, one can fit
a set of points \smash{$\left[\rho_i,E/A(\rho_i)\right]_{i=1,...,N}$}
which samples a realistic equation of state of nuclear matter.
Then, the parameters of the force will provide the characteristic
properties of nuclear matter ($\rho_0$, $E/A$, $K_\infty$ and
$m^*/m$) using equations (\ref{eq:effmass}) to (\ref{eq:pressure}).
In this kind of approach, the presence of a density-dependent
term with the power $2/3$ in the force is particularly
interesting. Indeed, the coefficient $\Theta_s$
is only constrained by the function $F(\Theta_s,t_{02})$ whose
value comes out of the fit. So, by choosing
correctly the parameter $t_{02}$ one can obtain any desired value of
the effective mass.

%
%

\section{RESULTS USING THE GENERALIZED SKYRME FORCES}

In this section, we present some results using the generalized
forces $[0,1,2]$, $[0,1,3]$ and $[0,2,3]$ for symmetric nuclear
matter as well as pure neutron matter. For this purpose, we need
to determine the coefficients $t_{0i}$, $x_{0i}$ and the two
combinations of coefficients $\Theta_s$ and
$\Theta_v=t_1(2+x_1)+t_2(2+x_2)$,
this latter expression is related to the isovector enhanced factor
$\kappa$~\cite{sly4} occurring in the Thomas-Reiche-Kuhn sum rule of the
isovector giant dipole resonance.
The quantities $E/A=-16$~MeV, $\rho_0=0.16$~fm$^{-3}$,
$K_\infty=230$~MeV, $m^*/m=0.8$ and $\kappa=0.5$ allow the determination
of the parameters $t_{0i}$, $\Theta_s$ and $\Theta_v$.
The coefficient $x_{0i}$ are determined by fitting the neutron
matter equation of state provided by Akmal {\it et al.}~\cite{akmal}.
The coefficients obtained accordingly are
summarized in Table~\ref{tab:a}.

\begin{table}[htb]
\caption{Coefficients $t_{0i}$ and $x_{0i}$ of the forces used
in this section. See the text for the remaining parameters.}
\label{tab:a}
\begin{center}
\begin{tabular}{|l|cc|cc|cc|cc|}
\hline
Force & $t_{00}$ & $x_{00}$ & $t_{01}$ & $x_{01}$ & $t_{02}$ & $x_{02}$
 & $t_{03}$ & $x_{03}$ \\
\hline
$[0,1,2]$
 & -1855.38 & 0.4006 & 2343.12 & 0.7589 & 
             \multicolumn{1}{r}{-488.28} & 3.6176 & -- & -- \\
$[0,1,3]$
 & -1807.41 & 0.3291 & 2078.04 & 0.4094 & -- & -- &
                       \multicolumn{1}{r}{-299.81} & 3.4464 \\
$[0,2,3]$
 & -1431.36 & 0.2828 & -- & -- & 3827.79 & 0.3814 & -2650.09 & 0.7640 \\
\hline
\end{tabular}
\end{center}
\end{table}

\begin{figure}[htb]
\begin{minipage}[t]{76mm}
\begin{center}
\includegraphics*[scale=0.35,angle=-90,bb= 114 88 539 694,clip]{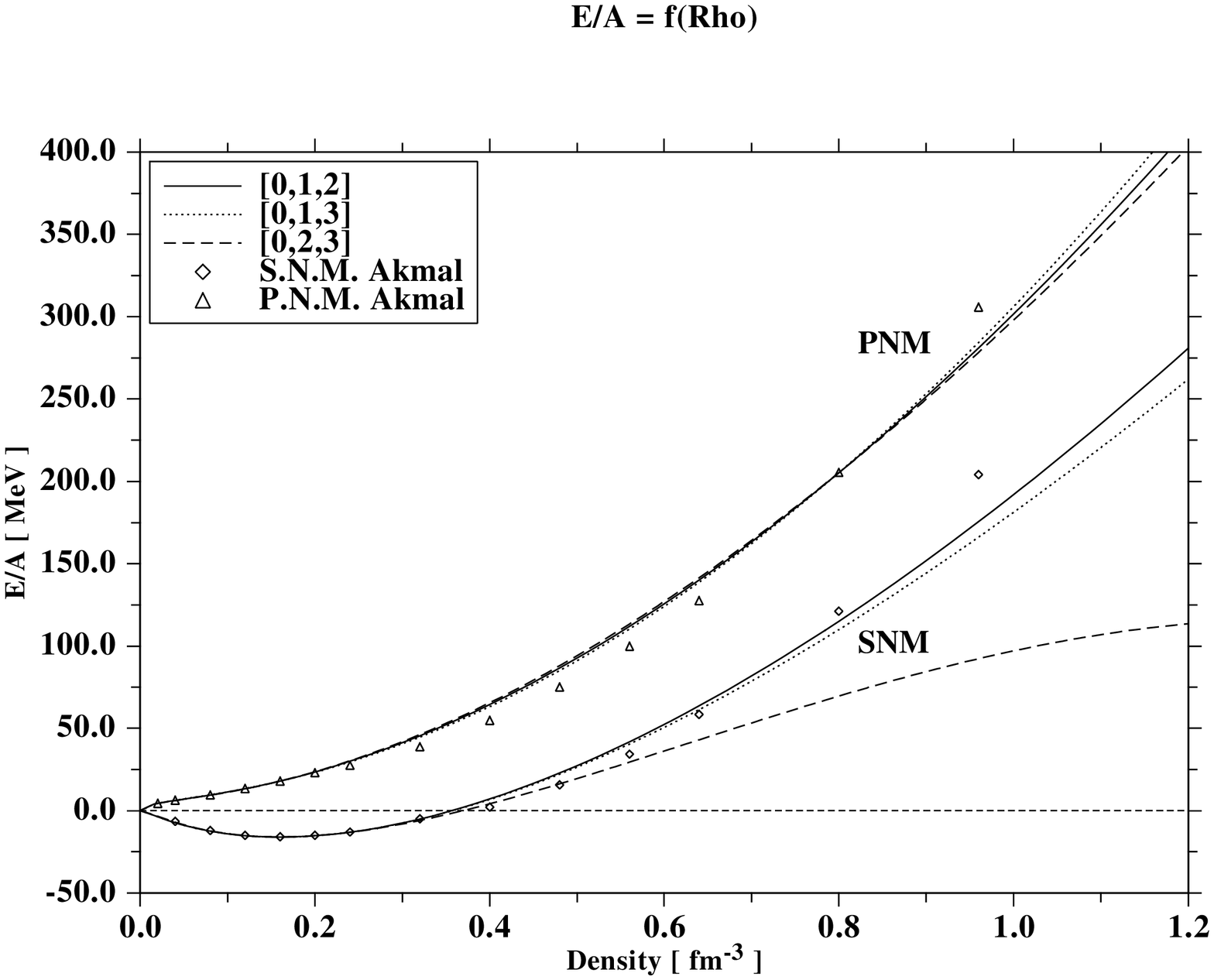}
\end{center}
\vskip -8mm
\caption{Energy per particle in infinite matter as a function
of equilibrium density.
SNM: symmetric nuclear matter; PNM: pure neutron matter. Labels
[0,1,2], [0,1,3], [0,2,3] refers to the density dependence
of the forces: see text.
Open diamonds and triangles: EOS of Akmal \etal~\cite{akmal}.}
\label{fig:eos}
\end{minipage}
\hspace{\fill}
\begin{minipage}[t]{76mm}
\begin{center}
\includegraphics*[scale=0.35,angle=-90,bb= 116 102 538 694,clip]{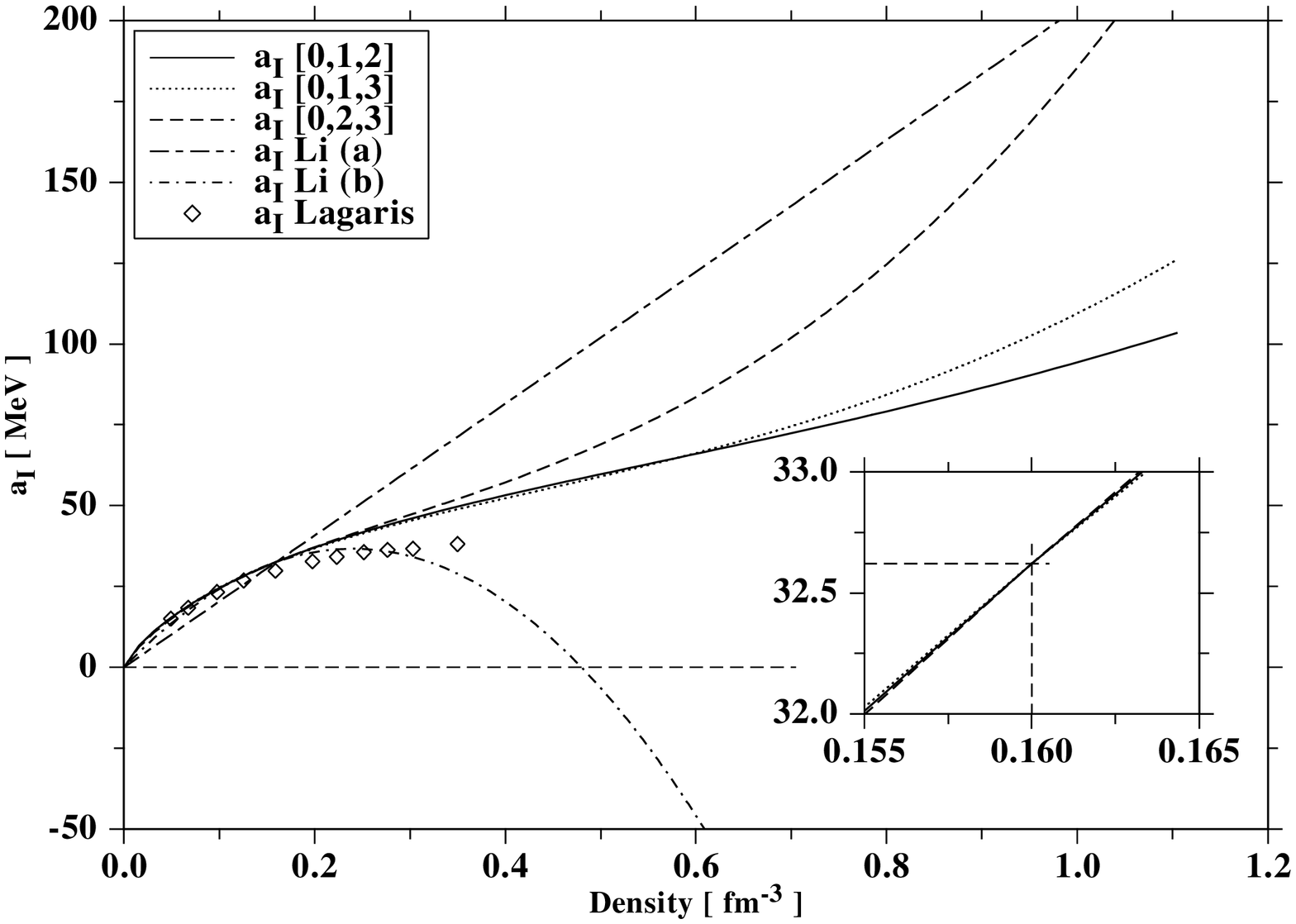}
\end{center}
\vskip -8mm
\caption{Symmetry energy in in\-fi\-nite matter as a function
of the density.
Li(a) and (b): phenomenological
parameterizations by Bao-An~Li~\cite{bao}, see text.
Open diamonds: variational calculation of Lagaris \etal~\cite{lagaris}.
Inset: symmetry energy in the vicinity of equilibrium density of symmetric
nuclear matter.}
\label{fig:aI}
\end{minipage}
\end{figure}

Figure~\ref{fig:eos} shows
the two EOS (energy per particle as a function of equilibrium density)
for symmetric nuclear matter and pure neutron matter
compared with the EOS of Akmal \etal~\cite{akmal}.
The results are quite reasonable and rather similar for
densities from 0 to  $\sim 3\rho_0$. The most salient feature
of the parameterizations is that the coefficient of the term
with the highest density dependence is always negative, leading
to the pathological property for high density
$E/A\longrightarrow -\infty$.
Nevertheless, as can be expected from the figure or check
numerically, this problem shows up only at extremely large and
unphysical densities and, thus, is not relevant
for the applications to atomic nuclei and neutron stars.

Another important property which can be extracted from this study
is the symmetry energy $a_I$. Its expression, obtained by a
straightforward calculation~\cite{sly4}, depends upon
the coefficients $t_{0i}$ and $x_{0i}$. On
Figure~\ref{fig:aI}, the symmetry energy is plotted as
a function of the density and compared with the variational results
of Lagaris \etal~\cite{lagaris} and the schematic parameterizations
of Bao-An~Li~\cite{bao}. The three forces [0,1,2], [0,1,3] and [0,2,3]
give results in good agreement with the variational
approach, especially for $\rho\lesssim0.3$~fm$^{-3}$, even though
this quantity has not been directly fitted.
For $\rho=\rho_0$, the three forces give exactly the same
symmetry energy, $a_I\simeq32.5$~MeV, which is close to the commonly
accepted value.
This result is very encouraging and reinforce the idea that
the two density-dependent terms in the Skyrme force open
pertinent degrees of freedom.

%
%

\section{CONCLUSION}

We have shown that the choice of a standard Skyrme effective force with
a modified density dependence based on two terms enables to choose
independently the incompressibility and the isoscalar effective mass.
The gross properties of nuclear matter investigated here are
quite reasonable: incompressibility ($\sim$ 230~MeV),
isoscalar effective mass ($\sim$ 0.8$\,m$) and symmetry energy
($\sim$ 32.5~MeV).
With this generalized density dependence we have been able to construct
new Skyrme like parameterizations without collapse at relevant densities
and which exhibit 
reasonable equation of state (EOS) for symmetric nuclear
matter as well as for pure neutron matter compared to the recent 
realistic variational EOS of Akmal~\etal~\cite{akmal}.

Other choices can be explored for the density dependence provided they
are physically well grounded.
The choice $\rho^{2/3}$ for one of the density-dependent term lets
total freedom for the effective mass. This feature is extremely
interesting from the perspective of developing accurate and
predictive force since it gives a control on the density of state
around the Fermi energy where the correlations beyond the Hartree-Fock
approximation can develop. Furthermore it implies only two
additional parameters so that their total number remains
quite small.

%
%

\newcommand{\APNY}[3]{Ann.~Phys.~(N.-Y.)~{#1}~(#3)~#2}
\newcommand{\NP}[3]{Nucl.~Phys.~{#1}~(#3)~#2}
\newcommand{\NPA}[3]{Nucl.~Phys.~A{#1}~(#3)~#2}
\newcommand{\PM}[3]{Philos.~Mag.~{#1}~(#3)~#2}
\newcommand{\PRep}[3]{Phys.~Rep.~{#1}~(#3)~#2}
\newcommand{\PRC}[3]{Phys.~Rev.~C{#1}~(#3)~#2}
\newcommand{\RMP}[3]{Rev.~Mod.~Phys.~{#1}~(#3)~#2}
\newcommand{\ibid}[3]{ibid.~{#1}~(#3)~#2}

\end{document}